\pdfoutput=1
\documentclass[aps,twocolumn,superscriptaddress,notitlepage, showpacs]{revtex4}
\usepackage[utf8]{inputenc}
\usepackage[T1]{fontenc}
\usepackage{lmodern}
\usepackage{amsmath,amssymb}
\usepackage{xcolor}
\usepackage{graphicx}
\usepackage[colorlinks,linkcolor=blue,urlcolor=blue,citecolor=blue]{hyperref}
\usepackage{mathptmx} 

\usepackage{xspace}

\newcommand{\eref}[1]{Eq.~(\ref{#1})}%

\newcommand{\erfc}{\mathrm{erfc}}

\begin{document}

\title{Exact distributions of cover times for $N$ independent random walkers in one dimension}

\author{Satya N. Majumdar}
\affiliation{LPTMS, CNRS, Univ. Paris-Sud, Universit\'e Paris-Saclay, 91405 Orsay, France}
\author{Sanjib Sabhapandit}
\affiliation{Raman Research Institute, Bangalore 560080, India}
\author{Gr\'egory Schehr}
\affiliation{LPTMS, CNRS, Univ. Paris-Sud, Universit\'e Paris-Saclay, 91405 Orsay, France}

\date{\today}

\begin{abstract}
We study the probability density function (PDF) of the cover time $t_c$ of a
finite interval of size $L$, by $N$ independent one-dimensional Brownian motions, each with
diffusion constant $D$. 
The cover time $t_c$ is the minimum time needed such that each point of the entire interval
is visited by at least one of the $N$ walkers. We derive exact results for the full PDF of $t_c$
for arbitrary $N \geq 1$, for both reflecting and periodic boundary conditions. The PDFs depend explicitly on $N$ and
on the boundary conditions. In the limit of large $N$,
we show that $t_c$ approaches its average value $\langle t_c \rangle \approx L^2/(16\, D \, \ln N)$, with fluctuations
vanishing as $1/(\ln N)^2$. We also compute the centered and scaled limiting distributions for large $N$ for both boundary conditions 
and show that they are given by nontrivial $N$-independent scaling functions.  
\end{abstract}


\maketitle

Stochastic search processes are ubiquitous in nature~\cite{Benichou:2011fy}. These include animals foraging 
for food \cite{Viswanathan:1996fr,Viswanathan:1999kf, Viswanathan-book}, various biochemical reactions \cite{Schlesinger,Luz:2009bb} such
as proteins searching for specific DNA sequences to bind \cite{Berg,Mirny,Gorman_review,Gorman_stepping} or sperm cells searching for an oocyte to fertilize \cite{eisenbach,Redner_Meerson}. Several of these stochastic search processes
are often modeled by a single searcher performing a simple random walk (RW) \cite{Benichou:2011fy,Luz:2009bb}. 
In many situations, the search takes place in a confined domain as the 
targets are typical scattered over the entire domain. Finding all these targets therefore requires an exhaustive exploration 
of this confined domain. In this context, an important observable that characterizes the efficiency of the search process is
the cover time $t_c$, i.e., the minimum time needed by the RW to visit all sites of this domain at least once \cite{Chupeau:2015gp}. The cover time
of a single random walker has also an important application in computer science, for instance for generating random spanning trees (with uniform measure) on an arbitrary connected and undirected graph $G$ \cite{Broder89,Aldous90}.

Computing analytically the statistics of $t_c$ for a given confined domain has remained
an outstanding challenge in RW theory. Most previous studies focused on calculating the mean
cover time on regular lattices, graphs and networks \cite{Aldous83, Broder_Karlin,Yokoi:1990vq,hilhorst,Hemmer:1998es,dembo,networks,ding}. Very recently, Chupeau {\it et al.} studied
the full distribution of the cover time on a finite graph by a {\it transient} RW, i.e., a walker that escapes to
infinity with a non-zero probability in the unbounded domain \cite{Chupeau:2015gp}. This includes in particular RWs on regular lattice
in dimensions $d>2$ (see also \cite{Belius}). For such transient RWs, Chupeau {\it et al.} found a rather robust result \cite{Chupeau:2015gp}, namely that the distribution 
of $t_c$, appropriately centered and scaled, approaches a Gumbel distribution, irrespective of the topology of the graph as well
as its boundary conditions. An important exception to this class of transient walkers is a RW in one or two dimensions, where
the walker is recurrent (i.e., starting from a given site, it comes back to it with probability one). It is thus natural to investigate the distribution
of $t_c$ for a RW in one or two dimensions. In particular, on a finite segment in $d=1$, is the scaled distribution of $t_c$ still given by a Gumbel law or is
it something completely different? This question is clearly relevant for any process modeled by a $1d$ RW in a finite domain, for instance for 
proteins searching for a binding site on a DNA strand \cite{Berg,Gorman_review}. Another important question concerns the role of the boundary conditions on the confined domain: how sensitive is the distribution of $t_c$ to the boundary conditions, in the limit of a large domain? In $d=1$, while the mean cover time $\langle t_c \rangle$ is known exactly for a RW on a finite interval of size $L$, for
both reflecting and periodic boundary conditions, computing the full distribution for these two boundary conditions has remained
an outstanding challenge. 
\begin{figure}
\includegraphics[width = \linewidth]{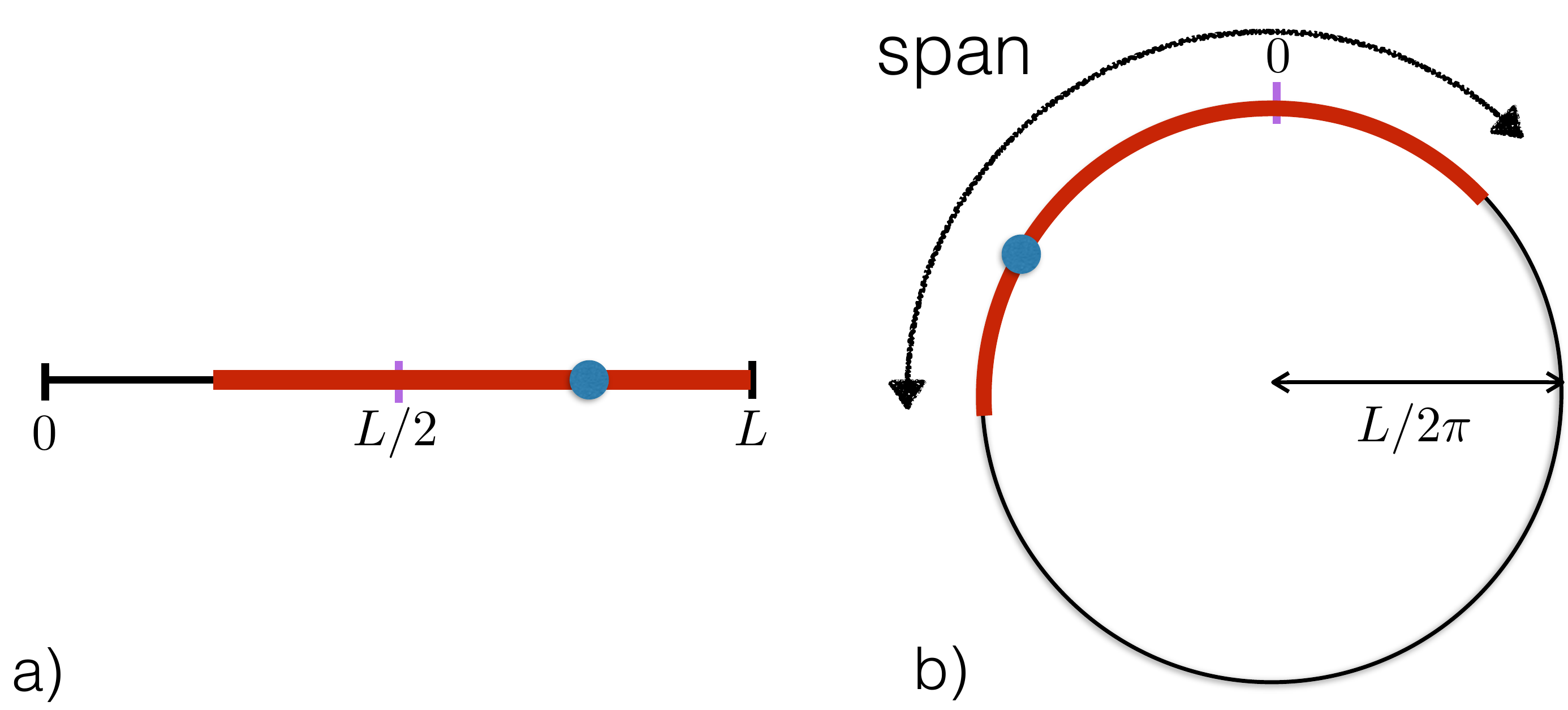}
\caption{(Color online) a) Brownian motion with reflecting boundary conditions (RBC) at $x=0$ and $x=L$. The thick (red) region indicates the space already covered by the walker up to time $t$, starting at $x_0 = L/2$. The circle denotes the current position, at time $t$, of the walker. b) The same walker on a ring, i.e., with periodic boundary conditions (PBC), starting at $0$. The thick (red) region, indicating the covered space up to time $t$, is equivalent to the span ${\cal S}(t)$ (the spatial extent of the visited region) of a walker on the infinite line. In both cases, the cover time $t_c$ is the first time at which the entire domain becomes~red.}\label{Fig1}
\end{figure}

In this Letter, we present exact results for the full distribution of $t_c$ in $d=1$ for a RW in the Brownian limit (i.e., the long time
scaling limit of a discrete-time RW on a lattice), on a finite interval of size $L$ for both reflecting (RBC) and periodic (PBC) boundary conditions~(see Fig. \ref{Fig1}). In the case of PBC, the RW takes place on a ring of size $L$ and evidently the distribution of $t_c$ is independent of the starting point, while it depends explicitly on the starting point $x_0 \in [0,L]$ in the case of RBC. In the latter case, for simplicity, we present the results only when the walker starts at the center of the interval, i.e., at $x_0=L/2$. We show that, in the Brownian limit (with a diffusion constant $D$), the probability density function (PDF) of $t_c$ is given by  
\begin{equation}
\mathrm{Prob.}[t_c =t|L]=\frac{4D}{L^2}\,
f_1^\mathrm{R|P}\left(\frac{4D\,t}{L^2}\right), 
\label{f1}
\end{equation}
where $\mathrm{R|P}$ denotes respectively the RBC and the PBC. The exact scaling functions $f_1^{\rm R}(x)$ and $f_1^{\rm P}(x)$ are given respectively in Eqs. (\ref{f1R}) and (\ref{f1P}), along with their asymptotics in Eqs. (\ref{asym-RBC}) and (\ref{asym-PBC}). Plots of these two scaling functions are shown in Fig. \ref{f1Rz}. 

Another interesting question concerns the statistics of the cover time $t_c$ where there are $N$ independent walkers.
This problem of multiple independent random walkers naturally arises in various search problems where there is a team of $N$ independent searchers, as
opposed to a single searcher. Various observables associated with this multiple random walker process have been studied over the last few decades, such as the first passage time to the origin \cite{Redner2001,our_review,oshanin,debacco}, the number of distinct and common sites visited by these walkers \cite{larralde:1992,Yuste,Tamm,Kundu:2013bk,Turban}, the statistics of the maximum displacement \cite{convex_hull_PRL,convex_hull_JSP,Paul,Anupam}, the statistics of records \cite{records},  etc. For $N$ walkers, the cover time $t_c$ is the minimum time needed for all sites to be visited at least once by at least one of the walkers. 
In the literature, only the mean cover time was computed and that too only for $N=2$ with PBC in $1d$ \cite{Hemmer:1998es}. It is evident that the average cover time will decrease with increasing $N$, but how does it decrease for large $N$? In this Letter, we generalize our result for the cover time distribution for one walker to arbitrary $N$ walkers in $1d$, both for RBC and PBC (for a plot of these distributions for different $N$, see Fig. \ref{fNRz}). In particular, we show that the mean cover time, for both boundary conditions, decreases for large $N$ as 
\begin{eqnarray}\label{mean_tc_N}
\frac{4 D \, \langle t_c \rangle }{L^2}  \approx \frac{1}{4 \ln N} \;.  
\end{eqnarray}
However, it turns out that the fluctuations around the mean are sensitive to the boundary conditions. Indeed we show that for large $N$, the random variable $t_c$ approaches to
\begin{eqnarray}\label{fluctuation_tc}
\frac{4 D \, t_c}{L^2}  \approx \frac{1}{4 \ln N} - \frac{1}{4 (\ln N)^2} \, \chi_{\rm R,P} \;,
\end{eqnarray}
where $\chi_{\rm R}$ and $\chi_{\rm P}$ are two $N$-independent distinct random variables with non-trivial PDFs given respectively by (with $x \in (-\infty, + \infty)$) 
\begin{eqnarray}\label{gR}
{\rm Prob.}[\chi_{\rm R} = x] = g_{\rm R}(x) = 2\,e^{-x-e^{-x}}\left(1-e^{-e^{-x}} \right)
\end{eqnarray}
plotted in Fig. \ref{fig_scaling}, and 
\begin{eqnarray}\label{gP}
{\rm Prob.}[\chi_{\rm P} = x] = g_{\rm P}(x) = 4\,e^{-2 x} \, K_0(2 \,e^{-x}) 
\end{eqnarray}
where $K_0$ is the modified Bessel function. The function $g_{\rm P}(x)$ is plotted in Fig. \ref{fig_scaling}.

\begin{figure}
\includegraphics[width=0.9\hsize]{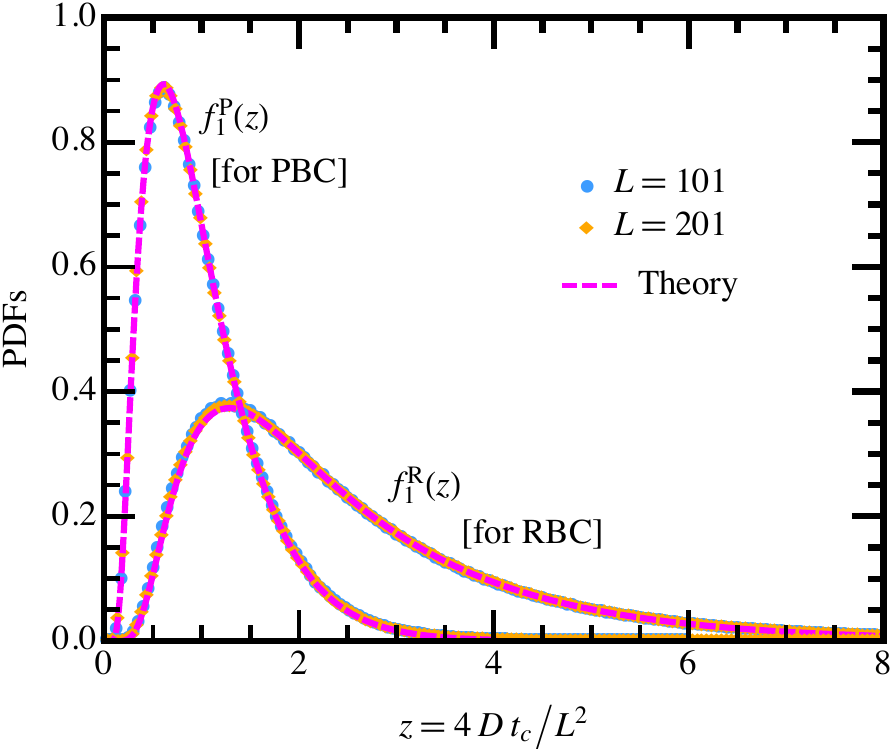}
\caption{\label{f1Rz}(Color online). The PDFs of the scaled cover time 
for a single lattice RW with RBC and PBC for sizes $L=101$ (blue) and $L=201$ (orange). The collapsed
scaling data are compared to theoretical scaling functions (dashed lines) in the Brownian limit  in 
Eq.~(\ref{f1}) with $D=1/2$, where $f_1^{\rm R|P}(z)$ are given respectively in Eqs. (\ref{f1R}) and (\ref{f1P}).}
\end{figure}

{\it Single walker (reflecting case)}.--- Let us start by first considering the case of a single
Brownian motion on the interval $[0,L]$ starting at $x_0$, with RBC at $x=0$ and
$x=L$. The cover time $t_c$ in this case is clearly the first time when the walker has hit both
boundaries at $x=0$ and $x=L$. It is useful to consider the cumulative distribution ${\rm Prob.}[t_c > t|x_0]$. 
If $t_c > t$, this means that at time $t$ one of the boundaries has not been hit up to time $t$. This means that 
${\rm Prob.}[t_c > t|x_0] = {\rm Prob.}[L \; {\rm \; is \; unhit \; up \; to \; time \;} t] + {\rm Prob.}[0 \;{\rm is \; unhit \; up \; to \; time \;} t] - 
{\rm Prob.}[{\rm both \; are \; unhit\; up \; to \; time \;}t]$. All the three probabilities can be computed by solving the standard backward Fokker-Planck
equation for the survival probability $S(x_0,t)$ ($x_0$ being the starting position of the walker)
\begin{equation}
\frac{\partial S(x_0,t)}{\partial t} = D\frac{\partial^2
S(x_0,t)}{\partial x_0^2} \; 
\label{FP-eqn}
\end{equation}
with appropriate boundary conditions at $x_0 = 0$ and $x_0=L$. For example, $ {\rm Prob.}[L \; {\rm \; is \; unhit \; up \; to \; time \;} t] = S_{\rm AR}(x_0,t)$ where the subscript ${\rm A}$ indicates an absorbing boundary condition at $x_0 = L$ (i.e., $S(x_0=L,t) = 0$), while the subscript ${\rm R}$ refers to the reflecting boundary condition at $x_0 = 0$, (i.e., $\partial_{x_0} S(x_0,t)|_{x_0=0} = 0$) \cite{Redner2001, satya_review,our_review}. Hence we have 
\begin{eqnarray}
{\rm Prob.}[t_c > t|x_0] = S_{\rm AR}(x_0,t) + S_{\rm RA}(x_0,t) - S_{\rm AA}(x_0,t) \;, \label{relation_prob}
\end{eqnarray}
where the subscripts refer to the boundary conditions. These survival probabilities can be computed exactly from Eq. (\ref{FP-eqn}) using standard methods \cite{Redner2001,our_review,chupeau_convex}, for instance either by expanding into Fourier modes (satisfying the appropriate boundary conditions) or equivalently by taking a Laplace transform with respect to time $t$ (for details, see \cite{SM}). For convenience, we will choose $x_0 = L/2$, for which by symmetry $S_{\rm AR}(x_0=L/2,t) = S_{\rm RA}(x_0=L/2,t)$. For this choice, one can show that
$S_\text{AA}(L/2,t)=S_1(4Dt/L^2)$ and
$S_\text{AR}(L/2,t)=S_\text{RA}(L/2,t)= S_2(4Dt/L^2)$, where
\begin{eqnarray}
\label{S1}
S_1(z)=\frac{4}{\pi}\sum_{n=0}^\infty  
\frac{(-1)^n}{(2n+1)}\,  
e^{-(2n+1)^2 \pi^2 z/4} \;,
\end{eqnarray}
and
\begin{eqnarray}
\label{S2}
S_2(z)~=\frac{4}{\pi}\sum_{n=0}^\infty 
\frac{(-1)^n\cos[(2n+1)\pi/4]}{(2n+1)}\,
e^{-(2n+1)^2 \pi^2 z/16} \;.
\end{eqnarray}
%
%\begin{subequations}
%\label{S1,2}
%\begin{align}
%\label{S1}
%%&S_1(z)=1-2\sum_{n=0}^\infty
%%(-1)^n\erfc\left(\frac{2n+1}{2\sqrt{z}}\right) \\
%%
%%\label{S1-2}
%&\quad\quad~
%S_1(z)=\frac{4}{\pi}\sum_{n=0}^\infty  
%\frac{(-1)^n}{(2n+1)}\,  
%e^{-(2n+1)^2 \pi^2 z/4},
%\intertext{and}
%\label{S2}
%%&S_2(z)=1-\sum_{n=0}^\infty
%%\left(\sin\frac{n\pi}{2}+\cos\frac{n\pi}{2} \right)
%%\erfc\left(\frac{2n+1}{2\sqrt{z}}\right)\\
%%\label{S2-2}
%S_2(z)~=\frac{4}{\pi}\sum_{n=0}^\infty 
%\frac{(-1)^n\cos[(2n+1)\pi/4]}{(2n+1)}\,
%e^{-(2n+1)^2 \pi^2 z/16} \;,
%\end{align}
%\end{subequations}
%where ${\rm erfc}(x) = (2/\sqrt{\pi})\,\int_{x}^\infty e^{-u^2} \, du$ is the complementary error function. 
%In Eqs. (\ref{S1}) and (\ref{S2}) the two different series representations are actually equivalent (related to each other
%via the Poisson summation formula). The first line is more suitable for the asymptotic analysis at small argument $z$ while the second
%line is more useful for the large $z$ analysis. 
Hence we have $\mathrm{Prob.}[t_c>t|L]
= F_1(4Dt/L^2)$, where the scaling function $F_1(z) = 2 S_2(z) - S_1(z)$. Taking derivative with respect to $z$
yields the PDF in Eq. (\ref{f1}) with the scaling function $f^{\rm R}_1(z)$ given explicitly by 
\begin{eqnarray}
f_1^{\rm R}(z) = S'_1(z) - 2\,S'_2(z) \label{f1R} \;,
\end{eqnarray}
where $S_1(z)$ and $S_2(z)$ are given in Eqs. (\ref{S1}) and (\ref{S2}). 
%
%\begin{subequations}
%\label{single RW reflecting}
%\begin{align}
%\label{f1z-1}
%&f_1^\mathrm{R}(z)= \frac{1}{\sqrt{\pi} z^{3/2}}\sum_{n=0}^\infty
%\left[\sin\frac{n\pi}{2}+\cos\frac{n\pi}{2} - (-1)^n\right]
% \notag\\
%&\qquad\qquad\qquad\qquad\qquad
%\times~  (2n+1)\,e^{-(2n+1)^2/(4 z)}
%\label{g1z-tail}
%&
%=\frac{\pi}{2}\sum_{n=0}^\infty
%%(-1)^n (2n+1)
%%\Bigl[\cos[(2n+1)\pi/4]\, e^{-(2n+1)^2\pi^2 z/16} \notag\\
%%&\qquad\qquad\qquad\qquad\qquad\quad
%% - 2 e^{-(2n+1)^2\pi^2 z/4} \Bigr],
%\end{align}
%\end{subequations}
The tails of the scaling function are given explicitly by (see Supp. Mat.~\cite{SM})
\begin{equation}
f_1^\mathrm{R}(z) \sim \begin{cases} (6/\sqrt{\pi}) z^{-3/2}
 e^{-9/(4z)} & \text{as}~ z\to 0,\\
\pi/\bigl(2\sqrt{2}\bigr) \,e^{-\pi^2 z/16}  &\text{as}~ z\to \infty.
\end{cases}
\label{asym-RBC}
\end{equation}
A plot of this scaling function is shown in Fig. \ref{f1Rz}, where it is also
compared to the simulation results. Simulations were done for a RW on a lattice of
$L=101$ and $L=201$ sites with reflecting boundary conditions, which in the long time
limit collapses to the Brownian scaling function in Eqs. (\ref{f1}) and~(\ref{f1R}).

{\it Single walker (periodic case)}.--- We now consider the cover time for a single RW on a
ring of length $L$. In this case, the distribution of $t_c$ is independent of $x_0$, which we take 
it to be at $0$ (see Fig. \ref{Fig1} b)). We first show that the cumulative probability ${\rm Prob.}[t_c>t|L]$ on the ring can be mapped exactly onto the cumulative distribution of the span ${\cal S}(t)$ of the walker 
at time $t$ on an infinite line -- the span being the length of the covered region by the walker up to time $t$. The probability that $t_c>t$ 
indicates that at time $t$ the ring has not been covered by the walker (see Fig. \ref{Fig1} b)). Since the ring has not been fully traversed at time $t$, the walker does not realize
that it is on a ring. Thus one can think of the walk taking place on an infinite line and ${\rm Prob.}[t_c>t|L]$ on the ring 
is just the probability that the span ${\cal S}(t)$ of the walker on the infinite line is less than $L$, i.e., one has the exact relation (see Fig. \ref{Fig1} b))
\begin{eqnarray}
{\rm Prob.}[t_c>t|L] = {\rm Prob.}[{\cal S}(t) < L] \label{mapping} \;.
\end{eqnarray}
The PDF of the span ${\cal S}(t)$ on the infinite line is known \cite{hughes_book}, ${\rm Prob.}[{\cal S}(t)=s] = (1/\sqrt{4Dt}) h_1(s/\sqrt{4 D t})$ where %the scaling function $h_1(y)$ is given by
\begin{eqnarray}\label{h1}
h_1(y) = \frac{8}{\sqrt{\pi}} \sum_{m=1}^\infty (-1)^{m+1} m^2 e^{-m^2\, y^2} \;.
\end{eqnarray}
Therefore, taking derivative of Eq. (\ref{mapping}) with respect to $t$, we obtain the PDF of the cover time on a ring as in Eq. (\ref{f1}) where the scaling function $f_1^{\rm P}(z) = 1/(2 \,z^{3/2})h_1(1/\sqrt{z})$. Using the explicit expression of $h_1(y)$ in Eq. (\ref{h1}) we then get
\begin{eqnarray}\label{f1P}
f_1^{\rm P}(z) = \frac{4}{\sqrt{\pi} z^{3/2}} \sum_{m=1}^\infty  (-1)^{m+1} m^2 e^{-m^2/z} \;.
\end{eqnarray}
The tails of this function are given by (see Supp. Mat. \cite{SM})
\begin{equation}
f_1^\mathrm{P}(z) \sim \begin{cases}
 \bigl(4/\sqrt{\pi}\bigr) z^{-3/2} e^{-1/z} & \text{as}~ z\to 0,\\
\pi^2 z \, e^{-\pi^2 z/4}  &\text{as}~ z\to \infty.
\end{cases}
\label{asym-PBC}
\end{equation}
For a plot of this scaling function, see Fig. \ref{f1Rz}.

%We now consider the cover time for a single RW on a
%ring containing $L$ sites. It is again convenient to first consider
%the events where the cover time is longer than $t$. Since the RW has
%not visited all the site by time $t$, we can cut open the ring at any
%one of the unvisited sites. Now it is easy to realize that the
%probability of the cover time on a ring is greater than $t$ is same as
%the probability for the span or the number of distinct sites visited
%is less than $L$ for a RW on the infinite line:
%\begin{math}
%\mathrm{Prob.}[t_c >t|L]=\mathrm{Prob.}[\mathrm{span} < L|t].
%\end{math}
%

\begin{figure}
\includegraphics[width=0.9\hsize]{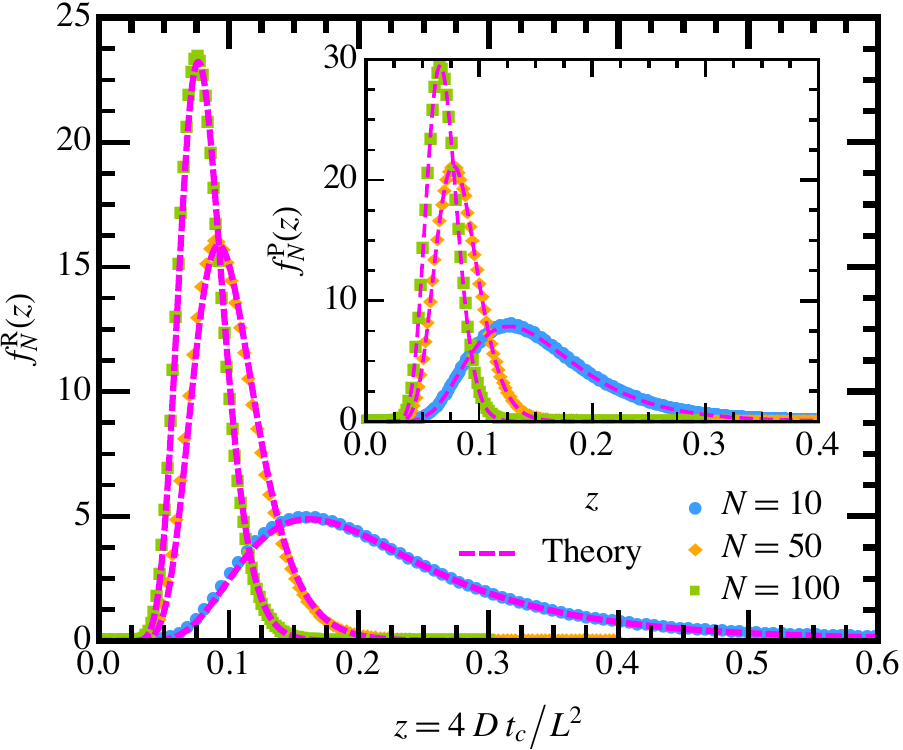}
\caption{\label{fNRz}(Color online). {\bf Main:} The PDFs of the scaled cover time (\ref{fN})
for different $N$ with RBC. For each $N$, the numerical results were obtained for lattice RW of
size $L=201$ as in Fig. \ref{f1Rz}. The
(magenta) dashed lines are the exact theoretical scaling functions in the Brownian limit with $D=1/2$, 
and are given by $f_N^{\rm R}(z)$ in Eq.~(\ref{scaled-f-RBC}). 
{\bf Inset:} Same data as shown in the main panel but for PBC. The (magenta) dashed lines correspond to the exact theoretical results given by $f_N^{\rm P}(z)$ in~(\ref{fNP_exact}).}
\end{figure}

{\it Multiple walkers (reflecting case).} --- Here we consider, for simplicity, $N$ independent walkers all
starting at the same point $x_0$. Using the mutual independence of the $N$ walkers, the cumulative cover time distribution for
$N$ walkers is clearly given by ${\rm Prob.}[t_c>t|x_0,N] = [S_\text{AR}(x_0,t)]^N + [S_\text{RA}(x_0,t)]^N - [S_\text{AA}(x_0,t)]^N$. Choosing as
before $x_0 = L/2$, we find that  
\begin{equation}
\mathrm{Prob.}[t_c =t|x_0=L/2, N]=\frac{4D}{L^2}\,
f_N^\mathrm{R}\left(\frac{4Dt}{L^2}\right), %\quad D=1/2,
\label{fN}
\end{equation}
with the superscript $\mathrm{R}$ denoting RBC and the scaling function given by
\begin{equation}
f_N^\mathrm{R}(z)=-F_N'(z) \;, \; {\rm where} \; \; F_N(z) = 2\bigl[S_2(z)\bigr]^N -\bigl[S_1(z)\bigr]^N  \label{scaled-f-RBC} \;,
\end{equation}
where $S_{1,2}(z)$ are given in Eqs. (\ref{S1}) and (\ref{S2}). A plot of this function for different values of $N$ is shown in the main panel of Fig. \ref{fNRz} where it is compared to simulations, with an excellent agreement. For $N \ge 2$, the asymptotics of $f_N^\mathrm{R}(z)$ are given by
\begin{equation}
f_N^\mathrm{R}(z) \sim \begin{cases}
2 N(N-1)/(\pi z)\, e^{-1/(2z)} & \text{as}~ z\to 0,\\
(N\pi^2/8) (2\sqrt{2}/\pi)^N\, e^{-N\pi^2z/16}  &\text{as}~ z\to \infty \;.
\end{cases}
\label{asym-RBC-N}
\end{equation}
Note that the small $z$ asymptotics of $f_N^{\rm R}(z)$ are quite different for $N=1$ (\ref{asym-RBC}) and $N \geq 2$ (\ref{asym-RBC-N}). 

One naturally wonders whether there exists a limiting distribution of $t_c$ for large $N$. We first estimate the mean cover time $4 D\langle t_c \rangle/L^2 = \int_0^\infty F_N(z) \, dz$, where $F_N(z)$ is the cumulative scaling function given in Eq. (\ref{scaled-f-RBC}). For large $N$, one expects that $\langle t_c \rangle$ is small. Hence, the integral $\int_0^\infty F_N(z) \, dz$ is dominated by the small $z$ behavior of $F_N(z)$. For small $z$, one can show from Eqs. (\ref{S1}) and (\ref{S2}) using Poisson summation formula (see \cite{SM} for details), that $S_1(z) \sim 1 - 4\sqrt{z/\pi} \, e^{-1/(4z)}$ and $S_2(z) \sim 1 - 2\sqrt{z/\pi} \, e^{-1/(4z)}$. Substituting this behavior in Eq. (\ref{scaled-f-RBC}) and exponentiating for large $N$ we get
\begin{eqnarray}
F_N(z) \approx 2\,e^{-u_N(z)} - e^{-2\,u_N(z)} \; , \; u_N(z) = \frac{2\sqrt{z}}{\sqrt{\pi}} N\, e^{-1/(4z)} \label{def_uN}  \;.\;
\end{eqnarray} 
Therefore $F_N(z) \sim 1$ as long as $u_N(z) \ll 1$ (which happens for $z < 1/(4 \ln N)$), while $F_N(z)$ is exponentially small in $N$ for
$z > 1/(4 \ln N)$. Hence, to leading order for large $N$, $4D\langle t_c \rangle/L^2 = \int_0^\infty F_N(z) \, dz \approx 1/(4 \ln N)$, as announced in Eq. (\ref{mean_tc_N}). In addition, we can also compute the limiting distribution from Eq. (\ref{def_uN}) by expanding around $z = 1/(4 \ln N)$. We set  
$z = 1/(4 \ln N) - x/(4 (\ln N)^2)$, where we assume that the scaled fluctuation $x$ is of order ${\cal O}(1)$. Substituting this $z$ in $u_N(z)$ in Eq. (\ref{def_uN}) and expanding for large $N$, one gets to leading order $u_N(z) \approx e^{-x}$. Hence, in this limit, one obtains $F_N(z) \to 2\,e^{-e^{-x}} - e^{-2\, e^{-x}}$. Taking derivative with respect to $x$ gives the limiting PDF of $t_c$ as announced in Eq. (\ref{gR}).

{\it Multiple walkers (periodic case)} --- We now consider $N$ independent walkers on a ring of size $L$,
all starting at the same point $0$. As in the $N=1$ case discussed earlier, the cumulative cover time
distribution is exactly related to the cumulative distribution of the span ${\cal S}_N(t)$ of $N$ walkers on
an infinite line, all starting at the same point, via the relation  
\begin{eqnarray}
{\rm Prob.}[t_c>t|L,N] = {\rm Prob.}[{\cal S}_N(t) < L] \label{mappingN} \;.
\end{eqnarray}
The study of the PDF of ${\cal S}_N(t)$ was initiated in Ref. \cite{larralde:1992} and was recently 
computed exactly for all $N$ in~\cite{Kundu:2013bk}. It was shown in Ref.~\cite{Kundu:2013bk} 
that  ${\rm Prob.}[{\cal S}_N(t)=s] = (1/\sqrt{4Dt}) h_N(s/\sqrt{4 D t})$ where the $N$-dependent scaling function
$h_N(y)$ is given by
\begin{eqnarray}
&&h_N(y)= \int_0^{\infty}dl_1 \int_0^\infty dl_2~\delta(y-l_1-l_2)
\frac{\partial^2g^N}{\partial l_1\partial l_2} \;. \label{span} 
%&& h_N(y)= \int_0^{\infty}dl_1 \int_0^\infty dl_2~\delta(y-l_1-l_2)
%\frac{\partial^2}{\partial l_1\partial l_2}\left[ g^N(l_1,l_2)\right] \;. 
\end{eqnarray}
Here $g(l_1,l_2)$ is the scaled cumulative joint distribution of the maximum and the minimum
of a single BM, starting at the origin on an infinite line and is given by \cite{Kundu:2013bk}: 
\begin{equation}
g(l_1,l_2)=\frac{4}{\pi} \sum_{n=0}^\infty \frac{1}{2n+1}~\text{{sin}}\left[\frac{(2\,n+1)\pi l_2}{{l_1}+l_2}\right]
e^{-\left[\frac{(2n+1)\pi}{2(l_1+l_2)}\right]^2}. 
\label{g}
\end{equation} 
As in the case of $N=1$, taking derivative of Eq. (\ref{mappingN}) with respect to $t$, we get 
\begin{equation}
\mathrm{Prob.}[t_c =t|L, N]=\frac{4D}{L^2}\,
f_N^\mathrm{P}\left(\frac{4Dt}{L^2}\right), %\quad D=1/2,
\label{fNP}
\end{equation}
with the superscript ${\rm P}$ denoting PBC. The scaling function $f_N^{\rm P}(z)$ is given by
\begin{eqnarray}\label{fNP_exact}
f_N^{\rm P}(z) = \frac{1}{2 \, z^{3/2}} \, h_N\left(\frac{1}{\sqrt{z}} \right) \;,
\end{eqnarray}
where $h_N(y)$ is given in Eq. (\ref{span}). In the inset of Fig. \ref{fNRz}, we show a plot
of $f_N^{\rm P}(z)$ for different $N$ and compare it to numerical results. The asymptotic tails
of $f_N^{\rm P}(z)$ for $N\geq 2$ are given by
\begin{equation}
f_N^\mathrm{P}(z) \sim \begin{cases}
\sqrt{2}N(N-1)/(\sqrt{\pi} z^{3/2})\, e^{-1/(2z)} & \text{as}~ z\to 0,\\
(a_N\, z/2)\, e^{-N \pi^2 z/4}  &\text{as}~ z\to \infty \;,
\end{cases}
\label{asym-PBC-N}
\end{equation}
where $a_N$ can be computed explicitly (see \cite{SM}). As in the reflecting case (\ref{asym-RBC-N}), the behavior
for $z \to 0$ is quite different for $N = 1$ and $N \geq 2$. 
%$a_N = 4 \pi^{3/2} N(N-1)~\left(4/\pi\right)^{N-2}~\Gamma([N-1]/2)/\Gamma(N/2)$

We now turn to the limiting distribution of $t_c$ for large $N$ for PBC. In the context of the span
distribution, the limiting form of the scaling function $h_N(y)$ was already analyzed for large $N$
in Ref. \cite{Kundu:2013bk} and it was found that 
\begin{eqnarray}\label{scaling_hN}
h_N(y) \approx 2 \, \sqrt{\ln N} \, {\cal D}\left( 2\,\sqrt{\ln N}\, (y - 2\,\sqrt{\ln N}) \right) \;,
\end{eqnarray}
where the function ${\cal D}(s) = 2 \,e^{-s}\, K_0(2 e^{-s/2})$ was obtained as a convolution of two 
Gumbel laws. Substituting this result (\ref{scaling_hN}) in Eq. (\ref{fNP_exact}) one finds that this function $f_N^{\rm P}(z)$ has
a sharp peak at $z = 1/(4 \ln N)$. To analyze the large $N$ scaling limit of $f_N^{\rm P}(z)$, we set $z = 1/(4 \ln N) - x/(4 (\ln N)^2)$ as
in the reflecting case. Expanding for large $N$, we get $f_N^{\rm P}(z) \approx 8 (\ln N)^2 {\cal D}(2\,x)$. Using $dz = dx/(4(\ln N)^2)$, one
immediately obtains the results for the periodic case announced in Eqs. (\ref{fluctuation_tc}) and (\ref{gP}). 

\begin{figure}
\includegraphics[width = 0.9\linewidth]{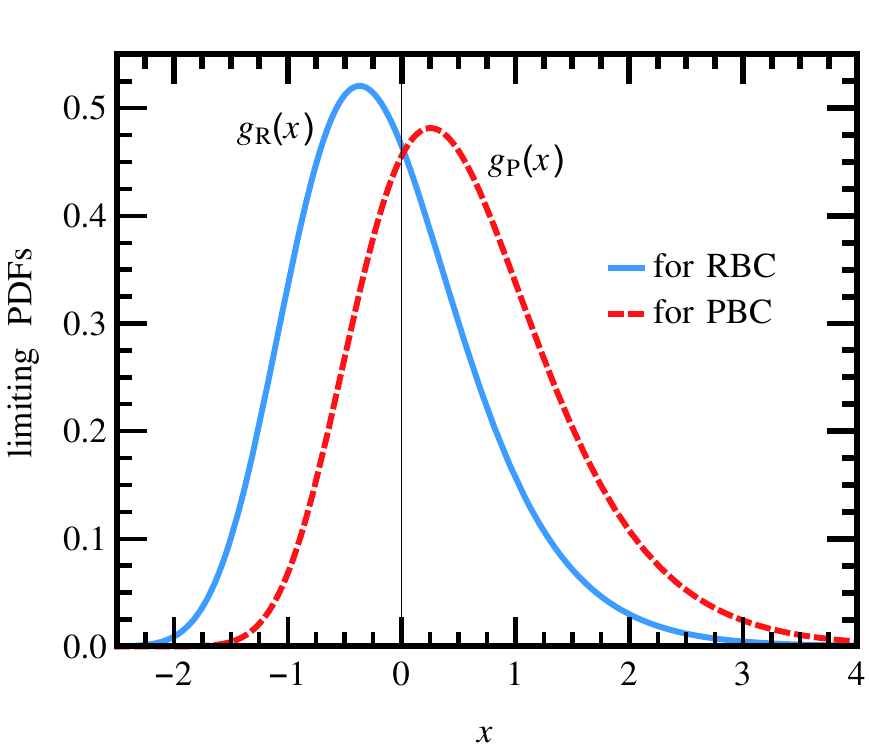}
\caption{Plot of the limiting PDFs of $t_c$ (for $N$ RWs in the limit of large $N$) $g_{\rm R}(x)$ (for RBC) and $g_{\rm P}(x)$ (for PBC) given respectively in Eqs. (\ref{gR}) and (\ref{gP}).}\label{fig_scaling}
\end{figure}

{\it Conclusion.}--- We have obtained the full PDF of the cover time $t_c$  
for $N$ independent Brownian motions in one dimension, both for reflecting and periodic
boundary conditions. Previously, only the first moment of $t_c$ was known in $1d$ for $N= 1$ and $N=2$. 
Our results provide the first instance of exact cover time distributions for
recurrent random walks, demonstrating clearly that this is different from a Gumbel law found recently 
for transient (i.e., non-recurrent) walks \cite{Chupeau:2015gp, Belius}. 
In addition, we have shown that in the limit of large $N$, the random
variable $t_c$ approaches its average value $\langle t_c\rangle \approx L^2/(16 \, D\, \ln N)$, with fluctuations
decaying as $1/(\ln N)^2$. The centered and scaled distributions converge to two distinct and nontrivial $N$-independent scaling functions $g_{\rm R}(x)$  and $g_{\rm P}(x)$ given respectively in Eqs. (\ref{gR}) and (\ref{gP}) and plotted in Fig. \ref{fig_scaling}. Another instance of recurrent RW is in $d=2$ for which the average value of $t_c$ has been well studied \cite{Aldous83,Broder_Karlin,hilhorst,dembo,ding}. However, determining its full PDF in $d=2$ for one or multiple ($N \geq 2$) walkers remains an outstanding challenge.

We thank the Indo-French Centre for the Promotion
of Advanced Research under Project Number 5604-E.

\newpage

\onecolumngrid

\begin{center}
{\Large Supplementary Material  \\ 
}
\end{center}

\begin{center}
We give some details about the results presented in the Letter not shown, for clarity, in the main text.  
\end{center}

\maketitle

%\abstract{We present some details of the calculations presented in the Letter.}

%\section{Backward Fokker-Planck equation for the survival probabilities}

{\it Reflecting boundary conditions (RBC).} We compute the survival probability $S(x_0,t)$ up to time $t$ 
of a $1d$ Brownian motion in $[0,L]$ starting at the initial position $x_0 \in [0,L]$. It satisfies the backward
Fokker-Planck equation \cite{our_review_app,Redner2001_app}
\begin{equation}
\frac{\partial S(x_0,t)}{\partial t} = D\frac{\partial^2
S(x_0,t)}{\partial x_0^2}, 
\label{FP-eqn}
\end{equation}
where $D$ is the diffusion constant. This equation holds for $x_0 \in [0,L]$, with the initial condition $S(x_0,0)=1$ for $0<x_0<L$. We consider three different boundary conditions, as needed in the text. For example we denote by $S_{\rm AR}(x_0,t)$ the survival probability corresponding to boundary conditions: absorbing (at $x_0=0$) and reflecting (at $x_0 = L$). Similarly, we will also compute $S_{\rm RA}(x_0,t)$ and $S_{\rm AA}(x_0,t)$. By taking Laplace transform $\tilde{S}(x_0,\lambda)=\int_0^\infty e^{-\lambda t} S(x_0,t)\, dt$
in \eref{FP-eqn}, and using the initial condition $S(x_0,0) = 1$, yields an ordinary differential equation
\begin{equation}
 D\frac{d^2\tilde{S}(x_0,\lambda)}{d x_0^2}
 -\lambda \tilde{S}(x_0,\lambda) =-1 \;.
\label{FP-eqn-Laplace}
\end{equation}
This differential equation can be trivially solved with the appropriate boundary conditions at $x_0=0$ and $x_0=L$. 
For simplicity, we choose $x_0=L/2$. In this case we obtain
\begin{align}
S_\text{AA}(L/2,t)=S_1\left(\frac{4Dt}{L^2}\right), \\
S_\text{AR}(L/2,t)=S_\text{RA}(L/2,t)=S_2\left(\frac{4Dt}{L^2}\right), 
\end{align}
where
\begin{align}
\int_0^\infty S_1(z) e^{-\lambda z}\,
dz&=\frac{1}{\lambda}\left[1-\frac{1}{\cosh\sqrt\lambda} \right] , \label{lap_S1} \\
\int_0^\infty S_2(z) e^{-\lambda z}\,
dz
&=\frac{1}{\lambda}\left[1-\frac{\cosh\sqrt\lambda}{\cosh2\sqrt\lambda} \right]. \label{lap_S2}
\end{align}
 
These Laplace transforms can be inverted by using the standard Bromwich contour in the complex $\lambda$-plane and
calculating the residues at the poles. This gives the results announced in Eqs. (8) and (9) in the text:
\begin{eqnarray}
\label{S1_supp}
S_1(z)=\frac{4}{\pi}\sum_{n=0}^\infty  
\frac{(-1)^n}{(2n+1)}\,  
e^{-(2n+1)^2 \pi^2 z/4} \;,
\end{eqnarray}
and
\begin{eqnarray}
\label{S2_supp}
S_2(z)~=\frac{4}{\pi}\sum_{n=0}^\infty 
\frac{(-1)^n\cos[(2n+1)\pi/4]}{(2n+1)}\,
e^{-(2n+1)^2 \pi^2 z/16} \; \;.
\end{eqnarray}
These series representations are very useful for calculating the large $z$ asymptotics, where only the $n=0$ term gives the
leading contribution. For example, from Eq. (10) in the text and the $n=0$ terms in $S_1(z)$ and $S_2(z)$ in 
Eqs. (\ref{S1_supp}) and (\ref{S2_supp}), it gives the large $z$ asymptotics of $f_1^{\rm R}(z)$ in Eq. (11) in the text. Similarly, for $N$ walkers, the large
$z$ asymptotics of the scaling function $f_N^{\rm R}(z)$ defined in Eq. (17) of the text can be derived from the leading $n=0$ term and this gives the second line of Eq. (18) of the text. 

However, for small $z$, it is harder to compute the tail from the series representations in (\ref{S1_supp}) and (\ref{S2_supp}). In that case, one can use an alternative 
representation that can be obtained via the Poisson summation formula \cite{Poisson}. Equivalently, we can derive it by inserting the following identity 
\begin{equation}
\frac{1}{\cosh(p\sqrt\lambda)}=2\sum_{n=0}^\infty (-1)^n
e^{-(2n+1)p\sqrt\lambda}, 
\end{equation}
in Eqs. (\ref{lap_S1}) and (\ref{lap_S2}) and then inverting the Laplace transform term by term. This gives, after straightforward algebra
\begin{align}
\label{S1_2_app}
S_1(z)&=1-2\sum_{n=0}^\infty
(-1)^n\erfc\left(\frac{2n+1}{2\sqrt{z}}\right), \\
\label{S2_2_app}
S_2(z)&=1-\sum_{n=0}^\infty
\left(\sin\frac{n\pi}{2}+\cos\frac{n\pi}{2} \right)
\erfc\left(\frac{2n+1}{2\sqrt{z}}\right) \;,
\end{align}
where $\erfc{(x)} = ({2}/{\sqrt{\pi}}) \int_{x}^\infty e^{-u^2} \, du$. Note that $\erfc{(x)} \approx e^{-x^2}/(x \sqrt{\pi})$ as $x \to \infty$. Consequently, for
small $z$, keeping only terms up to $n=1$ in the sums in Eqs. (\ref{S1_2_app}) and (\ref{S2_2_app}) (higher order terms only give subleading corrections), gives 
\begin{eqnarray}
&&S_1(z) \approx 1 - \frac{4 \sqrt{z}}{\sqrt{\pi}}\, e^{-\frac{1}{4z}} + \frac{4 \sqrt{z}}{3\sqrt{\pi}}\, e^{-\frac{9}{4z}} \;, \label{asympt_S1_app}\\
&&S_2(z) \approx 1 - \frac{2\sqrt{z}}{\sqrt{\pi}} \, e^{-\frac{1}{4z}} - \frac{2 \sqrt{z}}{3 \sqrt{\pi}} \, e^{-\frac{9}{4z}} \label{asympt_S2_app} \;.
\end{eqnarray}
Consequently, the scaling function for the cover time PDF, $f_1^{\rm R}(z) = S_1'(z) - 2 S_2'(z)$, has the asymptotics announced in Eq. (11) of the text. Note that, for small $z$, the leading term $\propto e^{-1/(4z)}$ actually cancels in $f_1^{\rm R}(z) = S_1'(z) - 2 S_2'(z)$. Similarly, for $N$ walkers ($N \geq 2$), using Eq. (17) for $f_N^{\rm R}(z)$ in the text and the above properties in (\ref{asympt_S1_app}) and (\ref{asympt_S2_app}) we get the small $z$ asymptotics in the first line of Eq. (18) in the text. Note that the leading small $z$ behavior of $f_N^{\rm R}(z)$ is rather different for $N=1$ and $N \geq 2$.

{\it Periodic boundary conditions (PBC).} We start with the scaling function $f_1^{\rm P}(z)$ given in Eq. (14) in the text:
\begin{eqnarray}
f_1^{\rm P}(z) = \frac{4}{\sqrt{\pi}z^{3/2}} \sum_{m=1}^\infty (-1)^{m+1} m^2 e^{-m^2/z} \;. \label{f1P_supp}
\end{eqnarray}
This formula is useful for the small $z$ asymptotics. Indeed, keeping the $m=1$ term in (\ref{f1P_supp}) gives the first line of Eq. (15) in the text.
However, this representation is not very convenient to derive the large $z$ asymptotics. For this, we could use the following identity:
\begin{eqnarray}\label{Poisson}
1 + 2 \sum_{m=1}^\infty (-1)^m e^{-m^2 x} = 2 \sqrt{\frac{\pi}{x}} \sum_{n=0}^\infty e^{- \frac{\pi^2}{x}(n+1/2)^2} \;, \;\;\;\; x > 0 \;,
\end{eqnarray}  
which can be easily derived from the Poisson summation formula \cite{Poisson}. Taking derivative with respect to $x$ on both sides and setting $x=1/z$ one obtains
\begin{eqnarray}\label{f1_Poisson}
f_1^{\rm P}(z) = \sum_{n=0}^\infty \left[  (2n+1)^2\, \pi^2 \, z- 2\right] e^{-\pi^2\,(n+1/2)^2\, z} \;.
\end{eqnarray}
For large $z$, the $n=0$ term provides the most dominant contribution, which gives the second line of Eq. (15) in the text. Using similar series representations for the function $g(l_1, l_2)$ defined in Eq. (22) in the text, and using results from Ref. \cite{Kundu_app} for the asymptotics of $h_N(y)$, one obtains the asymptotic results announced in Eq. (25) of the text with the coefficient
\begin{eqnarray}
a_N = 4 \pi^{3/2} N(N-1)~\left(4/\pi\right)^{N-2}~\Gamma([N-1]/2)/\Gamma(N/2) \label{aN} \;.
\end{eqnarray}
In particular one can check that $a_1 = 2 \pi^2$, in agreement with the second line of Eq. (15) in the text.


\begin{thebibliography}{9}




\bibitem{Benichou:2011fy} O. B\'enichou, C. Loverdo, M. Moreau, and
  R. Voituriez, Rev. Mod. Phys. {\bf 83}, 81 (2011).


\bibitem{Viswanathan:1996fr} G. M. Viswanathan, V. Afanasyev,
  S. V. Buldyrev, E. J. Murphy, P. A. Prince, and H. E. Stanley,
  Nature {\bf 381}, 413 (1996).

\bibitem{Viswanathan:1999kf} G. M. Viswanathan,
  S. V. Buldyrev, S. Havlin, M. G. E. da Luz, E. P. Raposo, and
  H. E. Stanley, Nature {\bf 401}, 911 (1999).


\bibitem{Viswanathan-book}
G. M. Viswanathan, M. G. E. da Luz, E. P. Raposo, and
H.E. Stanley, {\it The Physics of Foraging} (Cambridge Univ. Press,
Cambridge, 2011). 


\bibitem{Schlesinger}
M. F. Shlesinger, J. Phys. A: Math. Theor. {\bf 42}, 434001 (2009).




\bibitem{Luz:2009bb} For various stochastic search problems, see the special issue: {\it The Random Search Problem: Trends And Perspectives}, Eds. M. G. E. D. Luz, A. Grosberg, E. P. Raposo, and
G. M. Viswanathan,  J. Phys. A: Math. and Theor. {\bf 42}, 430301-434017 (2009).





\bibitem{Berg}
O. G. Berg, R. B. Winter, and P. H. Von Hippel, Biochemistry-US {\bf 20}, 6929 (1981).

\bibitem{Mirny}
L. Mirny, Nature Phys. {\bf 4}, 93 (2008).



\bibitem{Gorman_review}
J. Gorman, and E. C. Greene, Nat. Struct. Mol. Biol. {\bf 15}, 768 (2008). 


\bibitem{Gorman_stepping}
J. Gorman, A. J. Plys, M. L. Visnapuu, E. Alani, and E. C. Greene, Nat. Struct. Mol. Biol. {\bf 17}, 932 (2010). 


\bibitem{eisenbach}
M. Eisenbach, and L. C. Giojalas, Nat. Rev. Mol. Cell Biol. {\bf 7}, 276 (2006).


\bibitem{Redner_Meerson}
B. Meerson, and S. Redner, Phys. Rev. Lett. {\bf 114}, 198101 (2015).


\bibitem{Chupeau:2015gp}
M. Chupeau, O. B\'enichou, and R. Voituriez, Nat. Phys. {\bf 11}, 844
(2015).



\bibitem{Broder89}
A. Z. Broder, in {\it Proceedings of the Thirtieth Annual IEEE Symposium on Foundations of Computer Science}, 
(IEEE, New York, 1989), p. 442.

%{\it Generating random spanning trees}, in {\it Foundations of Computer Science} 442 (1990).


\bibitem{Aldous90}
D. J. Aldous, SIAM J. Discrete Math. {\bf 3},450 (1990).
%







\bibitem{Aldous83}
D. J. Aldous, Z. Wahrscheinlichkeit {\bf 62}, 36 (1983).
%On the time taken by random walks on nite groups to visit every state


\bibitem{Broder_Karlin}
A. Z. Broder, and A. R. Karlin, J. Theor. Proba. {\bf 2}, 101 (1989).


\bibitem{Yokoi:1990vq}
C. O. Yokoi, A. Hern\'andez-Machado, and L. Ram\'irez-Piscina, Phys.
Lett. A {\bf 145}, 82 (1990).

\bibitem{hilhorst}
M. J. A. M. Brummelhuis, and H. J .Hilhorst, Physica A {\bf 176}, 387 (1991).

\bibitem{Hemmer:1998es}
P. C. Hemmer, and S. Hemmer, Physica A {\bf 251}, 245 (1998).

\bibitem{dembo}
A. Dembo, Y. Peres, J. Rosen, and O. Zeitouni, Ann. Math. {\bf 160}, 433 (2004).



\bibitem{networks}
N. Zlatanov, L. Kocarev, Phys. Rev. E {\bf 80}, 041102 (2009).
%Rw on networks: pdf of cover time


\bibitem{ding}
J. Ding, Electron. J. Probab. {\bf 17}, 1 (2012).



\bibitem{Belius}
D. Belius, Probab. Theory Related Fields {\bf 157}, 635 (2013). 



\bibitem{Redner2001}
S. Redner, {\it A Guide to First-Passage Processes} (Cambridge
University Press, Cambridge, England, 2001).


\bibitem{our_review}
A. J. Bray, S. N. Majumdar, and G. Schehr, Adv. Phys. {\bf 62}, 225 (2013).


\bibitem{oshanin}
C. Mejia-Monasterio, G. Oshanin, and G. Schehr, J. Stat. Mech. P06022 (2011).

\bibitem{debacco}
U. Bhat, C. De Bacco, and S. Redner, J. Stat. Mech. P083401 (2016).


\bibitem{larralde:1992}
H. Larralde, P. Trunfio, S. Hamlin, H. E. Stanley, and G. H. Weiss, Nature (London) {\bf 355}, 423 (1992).


\bibitem{Yuste}
L. Acedo, and S. B. Yuste, Recent Res. Devel. Stat. Phys. {\bf 2}, 83 (2002). 
%Recent Research Developments in Statistical Physics, Volume 2, Pages 83-106 (Transworld Research Network, Trivandrum, India, 2002)


\bibitem{Tamm}
S. N. Majumdar, and M. Tamm, Phys. Rev. E {\bf 86}, 021135 (2012). 

\bibitem{Kundu:2013bk}
A. Kundu, S. N. Majumdar, and G. Schehr, Phys. Rev. Lett. {\bf 110},
220602 (2013).

\bibitem{Turban}
L. Turban, J. Phys. A {\bf 47}, 385004 (2014).



\bibitem{convex_hull_PRL}
J. Randon-Furling, S. N. Majumdar, and A. Comtet, Phys. Rev. Lett. {\bf 103}, 140602 (2009).


\bibitem{convex_hull_JSP}
S. N. Majumdar, A. Comtet, and J. Randon-Furling, J. Stat. Phys, {\bf 138}, 955 (2010). 


\bibitem{Paul}
P. L. Krapivsky, S. N. Majumdar, and A. Rosso, J. Phys. A: Math. Theor. {\bf 43}, 315001 (2010). 
%Comments: 18 pages, 9 figures
%Journal-ref: J. Phys. A: Math. Theor. 43 (2010) 315001





\bibitem{Anupam}
A. Kundu, S. N. Majumdar, and G. Schehr, J. Stat. Phys. {\bf 157}, 124 (2014). 


\bibitem{records}
G. Wergen, S. N. Majumdar, and G. Schehr, Phys. Rev. E {\bf 86}, 011119 (2012).  



\bibitem{hughes_book}
B. D. Hughes, {\it Random walks and random environments}, Clarendon Press Oxford, (1995), see formula (6.331) p. 384.  







\bibitem{SM} See Supplemental Material  for details. 


%\bibitem{Fisher1984} M. E. Fisher, J. Stat. Phys. {\bf 34}, 667 (1984).
%

\bibitem{satya_review}
S. N. Majumdar, Curr. Sci. {\bf 77}, 370 (1999).
 

\bibitem{chupeau_convex}
M. Chupeau, O. B\'enichou, and S. N. Majumdar, Phys. Rev. E {\bf 91}, 050104 (2015). 
%Convex hull of a Brownian motion in confinement

\end{thebibliography}

\begin{thebibliography}{9}




\bibitem{our_review_app}
A. J. Bray, S. N. Majumdar, and G. Schehr, Adv. Phys. {\bf 62}, 225 (2013).


\bibitem{Redner2001_app}
S. Redner, {\it A Guide to First-Passage Processes} (Cambridge
University Press, Cambridge, England, 2001).

\bibitem{Poisson}
See \url{https://en.wikipedia.org/wiki/Poisson_summation_formula}

\bibitem{Kundu_app}
A. Kundu, S. N. Majumdar, and G. Schehr, Phys. Rev. Lett. {\bf 110},
220602 (2013).

\end{thebibliography}
\end{document}